%% file: main.tex
\documentclass[10pt, conference, a4paper]{IEEEtran}
\IEEEoverridecommandlockouts
\input{preamble/header}
\input{acronyms/acronyms}

\usepackage{eso-pic}
\usepackage{url}
\AddToShipoutPictureBG*{
  \AtPageUpperLeft{%
    \put(0,-30){\raisebox{15pt}{\makebox[\paperwidth]{\begin{minipage}{21cm}\centering
      \textcolor{gray}{This work has been accepted for publication and presentation at the\\ 10th IEEE International Workshop on Advances in Sensors and Interfaces (IWASI 2025).} 
    \end{minipage}}}}%
  }
  \AtPageLowerLeft{%
    \raisebox{25pt}{\makebox[\paperwidth]{\begin{minipage}{21cm}\centering
      \textcolor{gray}{ \copyright 2025 IEEE.  Personal use of this material is permitted.  Permission from IEEE must be obtained for all other uses, in any current or future media, including reprinting/republishing this material for advertising or promotional purposes, creating new collective works, for resale or redistribution to servers or lists, or reuse of any copyrighted component of this work in other works.
      }
    \end{minipage}}}%
  }
}

\begin{document}
\input{content/00_01_title}
\input{content/00_02_abstract}

\input{content/01_intro}

\input{content/02_related_works}
\input{content/03_system_architecture}
\input{content/04_experimental_setup}
\input{content/05_results}
\input{content/06_conclusion}

\bibliographystyle{IEEEtran}
\bibliography{bib/IEEEabrv, bib/references}
\end{document}

%% file: preamble/header.tex
\usepackage[table]{xcolor}
\usepackage{multirow}
\usepackage{orcidlink}
\usepackage{booktabs}
\usepackage{cite}
\usepackage{amsmath,amssymb,amsfonts}
\usepackage[nolist,nohyperlinks]{acronym}
\usepackage[inline]{enumitem}
\usepackage[detect-weight=true, detect-family=true]{siunitx}
\usepackage{algpseudocode}
\usepackage{algorithm}
\usepackage{svg}
\usepackage{multirow}
\usepackage{balance}
\usepackage{graphicx}
\usepackage[nameinlink, capitalise]{cleveref}
\usepackage[usestackEOL]{stackengine}
\DeclareSIUnit\ohm{\ensuremath\Omega}
\DeclareSIUnit\db{dB}
\DeclareSIUnit{\belmilliwatt}{Bm}
\DeclareSIUnit{\belisotrope}{Bi}
\DeclareSIUnit{\bel}{B}
\DeclareSIUnit{\bitpersecond}{bps}
\DeclareSIUnit{\samplepersecond}{Sps}
\DeclareSIUnit{\nothing}{\relax}
\usepackage[percent]{overpic}
\usepackage[export]{adjustbox}

\usepackage{pifont}
\makeatletter
\newcommand*{\org@overidelabel}{}
\let\org@overridelabel\@verridelabel
\@ifpackagelater{acronym}{2015/03/21}{
  \renewcommand*{\@verridelabel}[1]{%
    \@bsphack
    \protected@write\@auxout{}{\string\AC@undonewlabel{#1@cref}}%
    \org@overridelabel{#1}%
    \@esphack
  }%
}{
  \renewcommand*{\@verridelabel}[1]{%
    \@bsphack
    \protected@write\@auxout{}{\string\undonewlabel{#1@cref}}%
    \org@overridelabel{#1}%
    \@esphack
  }%
}

\newcommand{\linebreakand}{%
  \end{@IEEEauthorhalign}
  \hfill\mbox{}\par
  \mbox{}\hfill\begin{@IEEEauthorhalign}
}
\makeatother

%% file: acronyms/acronyms.tex
\begin{acronym}
    \acro{A-TDOA}{asynchronous \ac{TDOA}}
    \acro{ADS-TWR}{alternative \acl{DS-TWR}}
    \acro{AP-TWR}{active-passive \ac{TWR}}
    \acro{AoA}{angle of arrival}
    \acro{BLE}{Bluetooth Low Energy}
    \acro{CC-SS-TWR}{\acl{CFO} corrected \acl{SS-TWR}}
    \acro{CFO}{carrier frequency offset}
    \acro{CIR}{channel-impulse response}
    \acro{DL-TDOA}{downlink \acl{TDOA}}
    \acro{DS-TWR}{double-sided \ac{TWR}}
    \acro{GPIO}{general purpose input-output}
    \acro{IoT}{Internet of Things}
    \acro{LNA}{low-noise amplifier}
    \acro{MCU}{microcontroller unit}
    \acro{OOK}{on-off keying}
    \acro{PCB}{printed circuit board}
    \acro{PLL}{phase-locked loop}
    \acro{RSSI}{received signal strength indicator}
    \acro{RTLS}{real-time locating system}
    \acro{RTT}{round-trip time}
    \acro{SPI}{serial peripheral interface}
    \acro{SS-TWR}{single-sided \acs{TWR}}
    \acro{TDMA}{time-division multiple-access}
    \acro{TDOA}{time difference of arrival}
    \acro{TOF}{time-of-flight}
    \acro{TWR}{two-way ranging}
    \acro{USB}{universal serial bus}
    \acro{UWB}{ultra-wideband}
    \acro{WuC}{wake-up call}
    \acro{WuR}{wake-up radio}
    \acro{UAA}{uniform angular array}
    \acro{ESA}{European Space Agency}
    \acro{ASIC}{application-specific integrated circuit}
    \acro{ASK}{amplitude-shift keying}
    \acro{FSK}{frequency-shift keying}
    \acro{RF}{radio frequency}
    \acro{LoRa}{Long Range}
    \acro{I2C}{inter-integrated circuit}
    \acro{NB-IoT}{narrowband \ac{IoT}}
    \acro{PGA}{programmable gain amplifier}
    \acro{IC}{integrated circuit}
    \acro{SDN}{shutdown}
    \acro{IRQ}{interrupt request}
    \acro{SMU}{source measure unit}
    \acro{PDR}{packet delivery ratio}
    \acro{LDR}{low data rate}
    \acro{HDR}{high data rate}
    \acro{mls}{maximum length sequence}
\end{acronym}

%% file: content/00_01_title.tex
\title{WakeMod: A 6.9\,\(\mathbf{\mu}\)W Wake-Up Radio Module with --72.6\,dBm Sensitivity for On-Demand IoT
}

\author{\IEEEauthorblockN{Lukas Schulthess\orcidlink{0000-0002-6027-2927}\IEEEauthorrefmark{1}, Silvano Cortesi\orcidlink{0000-0002-2642-0797}\IEEEauthorrefmark{1}, and  Michele Magno\orcidlink{0000-0003-0368-8923}}
\IEEEauthorblockA{\textit{Department of Information Technology and Electrical Engineering, ETH Zurich, Zurich, Switzerland}}
\IEEEauthorblockA{\IEEEauthorrefmark{1}Both authors contributed equally to the paper.}
}

\maketitle

%% file: content/00_02_abstract.tex
\begin{abstract}
Large-scale \ac{IoT} applications, such as asset tracking and remote sensing, demand multi-year battery lifetimes to minimize maintenance and operational costs.
Traditional wireless protocols often employ duty cycling, introducing a tradeoff between latency and idle consumption -- both unsuitable for event-driven and ultra-low power systems.

A promising approach to address these issues is the integration of always-on \acp{WuR}. They provide asynchronous, ultra-low power communication to overcome these constraints. 

This paper presents \textsc{WakeMod}, an open-source wake-up transceiver module for the \(\mathbf{\qty{868}{\mega\hertz}}\) ISM band. Designed for easy integration and ultra-low power consumption, it leverages the \(\mathbf{\qty{-75}{\deci\belmilliwatt}}\) sensitive \textsc{FH101RF} \ac{WuR}. 
\textsc{WakeMod} achieves a low idle power consumption of \(\mathbf{\qty{6.9}{\micro\watt}}\) while maintaining responsiveness with a sensitivity of \(\mathbf{\qty{-72.6}{\deci\belmilliwatt}}\). Reception of a wake-up call is possible from up to \(\mathbf{\qty{130}{\meter}}\) of distance with a \(\mathbf{\qty{-2.1}{\deci\belisotrope}}\) antenna, consuming \(\mathbf{\qty{17.7}{\micro\joule}}\) with a latency below \(\mathbf{\qty{54.3}{\milli\second}}\).
\textsc{WakeMod}'s capabilities have further been demonstrated in an e-ink price tag application, achieving \(\mathbf{\qty{7.17}{\micro\watt}}\) idle consumption and enabling an estimated 8-year battery life with daily updates on a standard \textsc{CR2032} coin cell. \textsc{WakeMod} offers a practical solution for energy-constrained, long-term \ac{IoT} deployments, requiring low-latency, and on-demand communication.
\end{abstract}

\acresetall

%% file: content/01_intro.tex
\section{Introduction}\label{sec:introduction}

The \ac{IoT} is progressively driven by small, battery-operated devices designed for durability and optimized for specialized tasks.
For large-scale applications such as asset tracking \cite{soori23_inter_thing_smart_factor_indus}, structural monitoring \cite{scharer24_towar_invas_monit_system_wind_turbin_blades}, and sensor deployments in harsh and remote environments \cite{schulthess24_passiv_async_wake_receiv_acous_under_commun}, a long battery lifetime is crucial to minimize costly maintenance.
Energy harvesting from existing sources, such as solar~\cite{nath23_inter_thing_integ_with_solar_energ_applic}, thermal~\cite{tuoi24_therm_energ_harves_using_ambien}, or \ac{RF}~\cite{makhetha24_integ_wirel_power_trans_low}, can contribute energy to prolong operation. However, such techniques do not influence the system's power needs. 
Whereas intelligent algorithms like context- and energy-aware sampling~\cite{bensaid24_energ_effic_adapt_sensin_framew, surrel20_event_trigg_sensin_high_qualit} can help to reduce the average power consumption by dynamically adjusting data generation, often devices still need to transmit and receive data from a central control unit.
However, while these solutions provide low-power configurations for reliable data transfer at variable data rates, they require periodic data exchange to maintain the connection state and synchronization with the gateway.
This makes the \ac{RF} frontend the most power-expensive subsystem, consuming significant power and limiting the devices' lifetime~\cite{muhoza23_power_consum_reduc_iot_devic, masoudi21_devic_vs_edge_comput_mobil_servic}.

To reduce the impact of the communication interface on the energy budget, \ac{RF} devices can be periodically deactivated, a method known as duty cycling~\cite{zimmerling21_synch_trans_low_power_wirel}. However, while duty cycling can be effective, it inevitably introduces latency and does not eliminate power consumption during idle listening~\cite{doudou13_survey_laten_issues_async_mac}.
In fact, most \ac{IoT} devices are event-based~\cite{sutton19_blitz} and do not need to send data periodically.
They should rather have a long battery lifetime, be responsive to external events, and operate reliably for years to reduce maintenance and operational costs~\cite{cortesi25_wakel}.
This is especially important for large-scale implementations, which require few, but if needed, real-time responses such as digital room reservation systems~\cite{szabo23_energ_timet_displ_meetin_rooms} or electronic price tags in supermarkets~\cite{e24_intel_integ_e_ink_displ}.
Ideally, such systems should consume no power in their idle state while still being responsive to external communication requests with low latency~\cite{doudou13_survey_laten_issues_async_mac}.
As a consequence, regular updates or synchronization events are not expedient, as they result in a tradeoff between latency and power consumption.

An alternative approach is the integration of always-on wireless receivers that have the ability to detect wireless messages of interest asynchronously, so-called \acp{WuC}, while consuming power in the micro- to nano-watt range~\cite{shellhammer23_wake_up_radio_concept, benbuk23_charg_wake_up_iot_devic} or even being fully passive~\cite{schulthess24_passiv_async_wake_receiv_acous_under_commun}. Typically, such a \ac{WuR} is combined with a traditional transceiver for highly efficient payload transfer~\cite{sutton19_blitz}.
Although this strategy minimizes transmission events, latency, and significantly reduces the power consumption, it does not eliminate it entirely~\cite{mercier22_low_power_rf_wake_up_receiv}. Special attention is therefore required during hardware design to optimize power consumption during the idle state.

This work presents and characterizes \textsc{WakeMod}, an open-source \qty{868}{\mega\hertz} wake-up transceiver module based on the ultra-low power \textsc{FH101RF} from \textsc{RFicicent} and \textsc{MAX41462} from \textsc{Analog Devices}.
It enables easy integration of wake-up capabilities into sensing devices and sensor networks by being a drop-in replacement for commercially available \textsc{HOPERF} sub-GHz modules such as the widely used \textsc{RFMx} modules.
Specifically, this article makes the following contributions:

\begin{enumerate}
\item The design and implementation of an open-source and easy-to-integrate ultra-low power wake-up transceiver module for sub-GHz communication (\textsc{WakeMod}).
\item A full characterization of the wake-up transceiver, including a detailed power profiling  during active and idle states, an analysis of the transmission power, receiver sensitivity, and \ac{PDR}.
\item The demonstration of \textsc{WakeMod}'s effectiveness in a real-world price tag application (\textsc{WakeTag}) to show the significance of a \ac{WuR} in large-scale applications.
\end{enumerate}

The rest of this article is organized as follows: \cref{sec:related_works} gives a summary of state-of-the-art \acp{WuR}. \cref{sec:system-architecture} details the design and implementation of \textsc{WakeMod} and \textsc{Wake-Tag}. The evaluation setup topology is outlined in \cref{sec:exp-setup}, with the corresponding measurement results presented in \cref{sec:results}. Finally, \cref{sec:conclusion} concludes this work.

%% file: content/02_related_works.tex
\section{Related Works}\label{sec:related_works}
Utilizing novel always-on ultra-low power \ac{WuR} has the potential to significantly decrease the energy consumption in devices that must remain in idle-listening mode.
This is particularly crucial in systems -- such as \acp{RTLS}~\cite{cortesi25_wakel} -- where low latency is essential.
Most solutions rely on \ac{OOK} modulation~\cite{bdiri17_tuned_rf_duty_cycled_wake, mikhaylov20_wake_ble, sultania20_enabl_low_laten_bluet_low}, as \ac{OOK} is particularly suited for low-power receivers due to the simplicity of decoding messages using periodic sampling.

\begin{table*}[htpb!]
    \vspace{-0.5cm}
    \caption{Overview of state-of-the-art \aclp{WuR}.}
    \label{tab:wur}
    \renewcommand{\arraystretch}{1.5}
    \input{tbl/related-works-soa}
    \vspace{-0.5cm}
\end{table*}

In~\cite{hambeck11} an ultra-low power \ac{ASIC} \ac{WuR} capable of sensing a wake-up pattern on \qty{868}{\mega\hertz} is presented. The receiver built by using an envelope detector, together with a low-noise baseband amplifier, \ac{PGA} and a mixed-signal correlation unit is capable of sensing signals with a power of \qty{-71}{\deci\belmilliwatt} whilst consuming only \qty{2.4}{\micro\watt} at \qty{1}{\volt}.

Spenza et al.~\cite{spenza15_beyon} introduced a \ac{WuR} operating at \qty{868}{\mega\hertz}, composed with commercially available components: a diode-based envelope detector, a comparator, and a \ac{MCU} managing the signal decoding. The \ac{WuR} attained a sensitivity of \qty{-55}{\deci\belmilliwatt} with a power consumption of \qty{1.276}{\micro\watt}. Utilizing a transmission power of \qty{10}{\deci\belmilliwatt}, they achieved a maximum wake-up range of \qty{45}{\meter}.

A comparable system utilizing off-the-shelf components has been proposed by Sutton et al.~\cite{sutton15_zippy}. The system comprises an \ac{OOK} transmitter utilizing the \textsc{TI CC110L}, while the receiver employs a passive \ac{OOK} demodulator succeeded by a \textsc{AS3930} \ac{WuR} for address decoding. The implemented system consumes \qty{8.1}{\micro\watt} during active listening, exhibiting a sensitivity of \qty{-51}{\deci\belmilliwatt}.

Polonelli et al.~\cite{polonelli21_ultra_low_power_wake_up} propose an ultra-low power \ac{WuR} architecture based on \ac{UWB} specifically targeting location-aware applications. Their approach utilizes \ac{OOK} modulation overlaid on the \ac{UWB} signal, simplifying the receiver design for low-power operation. The discrete component solution achieved a sensitivity of \qty{-48}{\deci\belmilliwatt} with a power consumption of \qty{100}{\micro\watt}. 

An \ac{ASIC} implementation realizing the protocol described by Polonelli et al. was presented by Villani et al. in~\cite{villani24_ultra_wideb_wake_up_receiv}. The chip complies with IEEE 802.15.4-2011, attaining a maximum sensitivity of \qty{-86}{\deci\belmilliwatt} with a wake-up latency of \qty{524}{\milli\second}, or \qty{-73}{\deci\belmilliwatt} with a latency of \qty{55}{\milli\second}. The power consumption is \qty{36}{\nano\watt} and \qty{93}{\nano\watt}, respectively. The wake-up range is not available due to the absence of in-field evaluations of the chip.

In~\cite{kazdaridis21_a_novel_archit}, Kazdaridis et al. introduce the \textsc{eWake} architecture, in the form of a semi-active \ac{WuR}. By employing a nano-power operation amplifier after the envelope detector, \textsc{eWake} aims to overcome some limitations of purely passive receivers. The authors report a sensitivity of \qty{-70}{\deci\belmilliwatt} whilst the power consumption could be kept below \qty{2}{\micro\watt}.

A final significant \ac{IC} is the \textsc{FH101RF} from \textsc{RFicient\textsuperscript{\textcopyright}}~\cite{frauenhofer_rficient}. The tri-band \ac{WuR} can receive \acp{WuC} at \qty{433}{\mega\hertz}, \qty{868}{\mega\hertz}, and \qty{2.4}{\giga\hertz} simultaneously. The chip achieves a sensitivity of \qty{-80}{\deci\belmilliwatt} and facilitates a configurable wake-up latency ranging from \qty{1}{\milli\second} to \qty{121}{\milli\second}, resulting in a power consumption between \qty{2.7}{\micro\watt} and \qty{87.3}{\micro\watt}.

The proposed \textsc{WakeMod} is designed around the \textsc{FH101RF} \ac{WuR}. Its commercial availability and state-of-the-art sensitivity make it an ideal choice. It outperforms similar solutions like the \textsc{STM32WL3x} from \textsc{STMicroelectronics}, which falls short in sensitivity and power efficiency for wake-up applications. Our work goes beyond simply comparing a \ac{WuR} chip: we present a fully integrated wake-up radio module, designed with practical deployment in mind. The module encapsulates all necessary subsystems for seamless operation, enabling plug-and-play integration into larger wireless sensor networks. Its compact design, standardized interfaces, and robustness make it easily scalable and suitable for real-world, large-scale network deployments.

%% file: tbl/related-works-soa.tex
\resizebox{\textwidth}{!}{
\begin{tabular}{@{}lrrrrrrrr@{}}
    \toprule
     & ISCAS '11\cite{hambeck11} & INFOCOM '15~\cite{spenza15_beyon} & SENSYS '15~\cite{sutton15_zippy} & WiMob '21~\cite{polonelli21_ultra_low_power_wake_up} & IPSN '21~\cite{kazdaridis21_a_novel_archit} & ISCAS '24~\cite{villani24_ultra_wideb_wake_up_receiv} & \textsc{RFicient}~\cite{frauenhofer_rficient}\\
      \midrule
      Modulation & \acs{OOK} & \acs{OOK} & \acs{OOK} & \acs{OOK} & \acs{OOK} & \acs{OOK} & \acs{OOK} \\
      Carrier freq. & \qty{868}{\mega\hertz} & \qty{868}{\mega\hertz} & \qty{434}{\mega\hertz} & \qty{3.99}{\giga\hertz} \acs{UWB} & \qty{868}{\mega\hertz} & \qty{3.99}{\giga\hertz} \acs{UWB} & \cellcolor{gray!25}tri-band\(^a\)\\
      Technology & \acs{ASIC} & Discrete & \textsc{AS3930} + Disc. & Discrete & Discrete & \acs{ASIC} & \cellcolor{gray!25}\textsc{FH101RF}\\
      Idle power & \qty{2.4}{\micro\watt} & \cellcolor{gray!25}\qty{1.28}{\micro\watt} & \qty{8.1}{\micro\watt} & \qty{100}{\micro\watt} & \qty{1.73}{\micro\watt} & \cellcolor{gray!25}36 -- \qty{93.2}{\nano\watt} & 2.7 -- \qty{87.3}{\micro\watt}\\
      Wake-up latency & \qty{40}{\milli\second} -- \qty{110}{\milli\second} & \qty{16}{\milli\second} & \qty{7.3}{\milli\second} & \qty{156.7}{\milli\second} & -- & 55 -- \qty{524}{\milli\second} & \multicolumn{1}{r}{\cellcolor{gray!25}1 -- \qty{121}{\milli\second}}\\
      RX sensitivity & -\qty{71}{\deci\belmilliwatt} & -\qty{55}{\deci\belmilliwatt} & -\qty{51}{\deci\belmilliwatt} & -\qty{48}{\deci\belmilliwatt} & \(<\)-\qty{70}{\deci\belmilliwatt} & \cellcolor{gray!25}-86 -- -\qty{73}{\deci\belmilliwatt} & -\qty{75}{\deci\belmilliwatt}\\
      Wake-up range & \cellcolor{gray!25} \qty{304}{\meter} & \qty{45}{\meter} & \qty{30}{\meter} & \qty{1}{\meter} & -- & -- & \qty{100}{\meter} \\
    \bottomrule
\end{tabular}
}\\[4pt]
\hspace*{0.1cm}\footnotesize{\(^a\)\qty{433}{\mega\hertz}, \qty{868}{\mega\hertz}, and \qty{2.4}{\giga\hertz}}

%% file: content/03_system_architecture.tex
\section{System Architecture}\label{sec:system-architecture}

\subsection{WakeMod}
\begin{figure}[h!b]
    \vspace{-0.3cm}
    \centering
    \begin{overpic}[width=0.9\columnwidth]{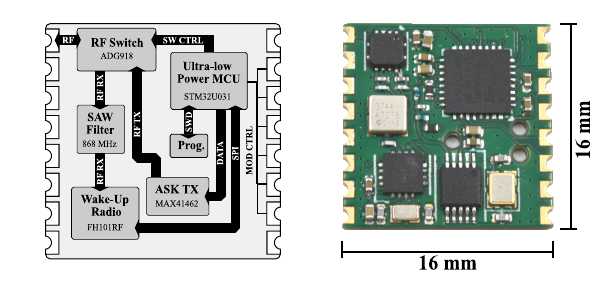}
        \put(1,5){(a)}
        \put(50,5){(b)}
    \end{overpic}
    \vspace{-5mm}
    \caption{System overview: High-level block diagram of the main platform \textsc{WakeMod} (a) and its hardware implementation (b).}
    \label{fig:wakemod}
    \vspace{-0.3cm}
\end{figure}

The designed \textsc{WakeMod}, as shown in~\cref{fig:wakemod}, is a \qty{868.35}{\mega\hertz} \ac{WuR} transceiver, built around a careful selection of components aiming for an ultra-low power consumption.
At its core is the \textsc{RFicient FH101RF}, serving as the \ac{WuR} of the module. The \textsc{FH101RF} is an ultra-low power tri-band (\qty{433}{\mega\hertz}, 868/\qty{915}{\mega\hertz}, \qty{2.4}{\giga\hertz}) \ac{WuR}, capable of receiving \ac{OOK}-modulated messages with a sensitivity of \qty{-75}{\deci\belmilliwatt}.
After recognizing a one-bit preamble, the \ac{WuR} can be addressed with a \qty{16}{\bit} wake-up address. 
The \ac{OOK}-messages for both preamble and address are encoded using a \qty{32}{\bit} \ac{mls}, configured individually with a data rate between \qty{256}{\bit\per\second} to \qty{32.768}{\kilo\bit\per\second}. This allows for a low-power consumption down to \qty{4.7}{\micro\watt} during idle listening of the \ac{WuR}, while still being able to achieve a wake-up latency below \qty{150}{\milli\second}.
Additionally, a custom \qty{6}{\byte} payload message can be received and decoded by the \ac{WuR}\footnote{The datasheet wrongly states a buffer of \qty{40}{\bit}, although it can hold \qty{48}{\bit}.}.

To handle \ac{WuC} transmission, the module incorporates the \textsc{Analog Devices MAX41462} \ac{OOK} transmitter, pre-configured for the \qty{868.35}{\mega\hertz} band. The transmitter remains in shutdown state until a toggle on its data line triggers it to activate and modulate the incoming data using \ac{OOK}.
Additionally, the \textsc{MAX41462} has an auto-shutdown feature that starts when data transmission stops, significantly lowering the transmitter's power consumption to \qty{34.2}{\nano\watt}.
However, on data rates below \qty{1024}{\bitpersecond}, sequences of '0's can trigger premature auto-shutdown, leading to message corruption due to the slow turn-on time.

To eliminate the need for connecting two antennas to \textsc{WakeMod}, thus reducing space requirements, cost, and complexity, the \ac{WuR} and the transmitter are connected to a single antenna through an \textsc{Analog Devices ADG918} \ac{RF} switch.

The module's control logic is implemented using an \textsc{STMicroelectronics STM32U031} \ac{MCU}. This specific \ac{MCU} was chosen for its ultra-low shutdown consumption of only \qty{29}{\nano\watt} together with its fast wake-up time of \qty{290}{\micro\second}, making it ideal for ultra-low power or even energy-harvesting based applications.
Its \ac{I2C} slave capabilities are used to make the module controllable by an external host \ac{MCU} to configure the \ac{WuR} for a specific wake-up address, read out wake-up data, or adjust settings such as data rates or reception branch.

Having a similar \qtyproduct{16x16}{\milli\meter} footprint as the \textsc{HopeRF RFMx} modules, \textsc{WakeMod} can serve as a drop-in replacement for existing designs, allowing effortless retrofit of \ac{WuR} capabilities into existing solutions.
Furthermore, the design of \textsc{WakeMod} is published as open source on \textsc{GitHub}\footnote{\url{https://github.com/ETH-PBL/WakeMod}}, making it readily available to developers. 

The firmware architecture of the module is optimized for ultra-low power consumption, keeping the \textsc{STM32U031} as long as possible in shutdown mode. Doing so requires keeping critical data in the backup registers of the tamper block.
Upon waking up, the \ac{MCU} identifies the source of the wake-up event, which can be one of the following:

\begin{description}
    \item[System Reset:] In the event of a system reset, the \ac{MCU} transitions immediately back to shutdown, ensuring minimal power consumption.
    \item[\Ac{WuR} \ac{IRQ}:] On a wake-up triggered by the \ac{WuR}, the modules clears the interrupt and retrieves the interrupt source with up to \qty{6}{\byte} of payload data from the \ac{WuR} FIFO. This information is stored in the \ac{MCU}'s backup register, before the host \ac{MCU}'s \ac{IRQ} is asserted and the module returns to shutdown.
    \item[\Ac{SDN} pin:] If the \ac{SDN} pin transitions to low, the \ac{MCU} wakes up and prepares itself to receive commands from the host over its \ac{I2C} slave interface. Four commands are supported:
    \begin{enumerate*}[label=(\roman*),,font=\itshape]
      \item \texttt{WhoAmI} for device identification; 
      \item \texttt{SetupWuR}, configuring the \ac{WuR} with a wake-up address, low data-rate, and high data-rate. This includes a full calibration of the \textsc{FH101RF};
      \item \texttt{SendWuC}, sending a \ac{WuC} to a specific wake-up address via the \textsc{MAX41462}, utilizing \ac{SPI} for accurate timing of preamble and address/payload transmission;
      \item \texttt{IRQReason}, retrieving the last wake-up event's reason and payload from backup registers.
    \end{enumerate*}
\end{description}



\subsection{WakeTag}
\begin{figure}[htpb!]
    \vspace{-0.3cm}
    \centering
    \begin{overpic}[width=0.85\columnwidth]{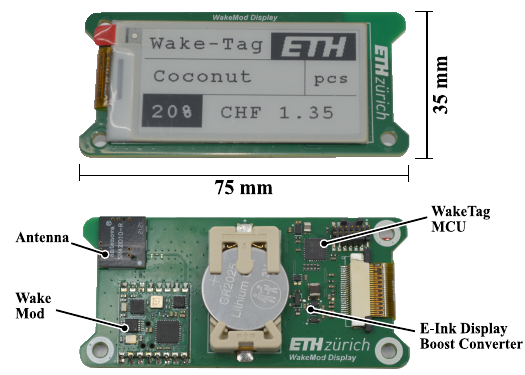}
        \put(1,42){(a)}
        \put(1,2){(b)}
    \end{overpic}
    \vspace{-3mm}
    \caption{WakeTag: (a) Front view, (b) Back view.}
    \label{fig:waketag}
    \vspace{-4mm}
\end{figure}

To test \textsc{WakeMod} in a real-world application scenario and demonstrate the power-saving potential of asynchronous wireless updates using a wake-up receiver, a custom-designed e-ink display system named \textsc{WakeTag} has been developed (see~\cref{fig:waketag}).

The ultra-low power Arm Cortex-M0+ \ac{MCU} \textsc{STM32U031} is the heart of the \textsc{WakeTag}, handling the \textsc{WakeMod} initialization and controlling the single-colored 2.13-inch e-ink display with \qtyproduct{250x122}{} pixels of type \textsc{ZJY122250-0213BBDMFGN-R}. 
As the display needs at least \qty{2.7}{\volt}, it is directly powered by the non-rechargeable CR2032-sized \qty{3}{\volt} coin cell battery. A dedicated boost converter circuit generates the positive and negative voltages required to drive the pixel electrodes. 
Both sub-systems are powered through a \textsc{Vishay SIP32431} power switch and can be disconnected from the battery to minimize power consumption.
To achieve the most efficient operating point for the \textsc{WakeMod} module, the digital communication over \ac{I2C} and power supply of the display's microcontroller are set to \qty{1.8}{\volt}, which is provided by the \textsc{TI TPS62840} high-efficiency step-down converter.
The \ac{SPI} communication to the display is established through its internal level shifter, which can be separately powered from the VDDIO by the system voltage and is directly powered via a \ac{GPIO} from the microcontroller.
Thus, the display can be fully decoupled from the power, eliminating its static power consumption.
At the same time, the information remains visible due to the bistable nature of its electrophoretic particles, which only require power for state changes.


The \textsc{WakeTag} firmware is kept very simple, prioritizing ultra-low power operation. Upon reset, its first action is to configure the \ac{WuR} using the \texttt{SetupWuR}.
Once configured, the \ac{MCU} enters shutdown mode to conserve energy.
On an \ac{IRQ} signal received from the \textsc{WakeMod}, the firmware reads the \ac{IRQ} reason.
If the \ac{IRQ} is caused by a \ac{WuR} message, the \qty{6}{\byte} payload is retrieved.
This payload contains the information for the price tag application: item, price, discount, and marking if it's out of stock.
The display is then activated and refreshed with this data before the system enters shutdown mode again.

%% file: content/04_experimental_setup.tex
\section{Experimental Setup}\label{sec:exp-setup}
\begin{table*}
\vspace{-0.5cm}
    \begin{center}
    \caption{Power consumption and TX power of \textsc{WakeMod} during listening and transmitting}
    \label{tab:pwr}
    \vspace{-0.5cm}
    \resizebox{\textwidth}{!}{
    \renewcommand{\arraystretch}{1.5}
    \begin{tabular}{@{}lr@{}}
        \multicolumn{2}{c}{\textsc{a) Idle listening consumption}}\\
        \toprule
        Data rate & Power consumption\\
        \midrule
        \qty{1024}{\bitpersecond} & \qty{6.88}{\micro\watt} \\
        \qty{2048}{\bitpersecond} & \qty{10.08}{\micro\watt}\\
        \qty{4096}{\bitpersecond} & \qty{16.54}{\micro\watt}\\
        \qty{8192}{\bitpersecond} & \qty{29.41}{\micro\watt}\\
        \qty{16384}{\bitpersecond} &\qty{55.01}{\micro\watt}\\
        \qty{32768}{\bitpersecond} & \qty{105.88}{\micro\watt}\\
        \bottomrule
    \end{tabular}
    \quad
    \begin{tabular}{@{}llr@{}}
        \multicolumn{3}{c}{\textsc{b) TX power and consumption}}\\
        \toprule
        Voltage & TX Power & Power consumption\\
        \midrule
        \qty{1.8}{\volt} & \qty{2.78}{\deci\belmilliwatt} & \qty{26.08}{\milli\watt}\\
        \qty{2.0}{\volt} & \qty{4.98}{\deci\belmilliwatt} & \qty{34.46}{\milli\watt}\\
        \qty{2.5}{\volt} & \qty{8.32}{\deci\belmilliwatt} & \qty{58.68}{\milli\watt}\\
        \qty{2.75}{\volt} & \qty{9.31}{\deci\belmilliwatt} & \qty{73.04}{\milli\watt}\\
        \qty{3.0}{\volt} & \qty{10.10}{\deci\belmilliwatt} & \qty{88.44}{\milli\watt}\\
        \qty{3.3}{\volt} & \qty{10.92}{\deci\belmilliwatt} & \qty{108.54}{\milli\watt}\\
        \bottomrule
    \end{tabular}
    \quad
    \begin{tabular}{@{}lrr@{}}
        \multicolumn{3}{c}{\textsc{c) Consumption of auxiliary operations}}\\
        \toprule
         & Energy consumption & Duration\\
        \midrule
        \textsc{WakeMod} \texttt{WhoAmI} & \qty{26.59}{\micro\joule} & \qty{15.9}{\milli\second} \\
        \textsc{WakeMod} \texttt{SetupWuR}\(^a\) & \qty{1.14}{\milli\joule} & \qty{564.2}{\milli\second} \\
        \textsc{WakeMod} \texttt{SendWuC}\(^b\) & \qty{106.15}{\micro\joule}\(^b\) & \qty{25.7}{\milli\second}\(^b\)\\
        \textsc{WakeMod} \texttt{IRQReason} & \qty{57.54}{\micro\joule} & \qty{18.9}{\milli\second}\\
        \acs{IRQ} handling (no payload) & \qty{15.88}{\micro\joule} & \qty{7.4}{\milli\second} \\
        \acs{IRQ} handling (\qty{6}{\byte} payload) & \qty{46.64}{\micro\joule} & \qty{19.6}{\milli\second} \\
        \bottomrule
    \end{tabular}
    }
    \end{center}
    \hspace*{0.2cm}{\footnotesize\(^a\) includes a full calibration of the \textsc{FH101RF}.}\\
    \hspace*{0.2cm}{\footnotesize\(^b\) cost of the actual transaction must be added with TX power consumption multiplied by transmission duration.}
    \vspace{-0.5cm}
\end{table*}

The characterization of \textsc{WakeMod} has been conducted based on power consumption, communication range, spectral emission, and output power. Finally, \textsc{WakeMod} has been tested within the \textsc{WakeTag} application to demonstrate its functionality in a real-world application.

\subsection{Power Consumption Measurements}
Precise measurements were conducted using a \textsc{Keysight N6705C} DC Power Analyzer fitted with a \textsc{Keysight N6781A} \ac{SMU}. The module's current draw has been characterized under the following conditions:
\begin{description}
    \item[Idle/Sleep Mode:] Measuring the baseline power consumption while the module awaits a \ac{WuC} at different data rates. Thereby, the \ac{LDR} is the data rate of the preamble's \ac{mls}, and the \ac{HDR} of the address and payload's \ac{mls}.
    \item[Active Reception:] Measuring the power consumption while the \ac{WuR} is actively listening for and processing an incoming \ac{WuC}.
    \item[Active Transmission:] Measuring the power consumption while the \textsc{MAX41462} transmitter actively sends a \ac{WuC}.
    \item[Configuration Mode:] Characterizing transient power draw during module initialization, whoami request, or \ac{IRQ} readout via the host \ac{MCU}.
\end{description}
For the active transmission and reception states, the characterization is performed across different configurations, including different data rates and varying payload sizes, to assess their impact on the overall energy consumption per event.

\subsection{Transmitter and Receiver Characterization}
An \textsc{R\&S SMBV100A} signal generator was used to characterize \textsc{WakeMod}'s sensitivity, configured to output an \ac{OOK} signal on an \qty{868.35}{\mega\hertz} carrier. This signal employed a \qty{1024}{\bitpersecond} data rate for the \ac{mls} of the preamble and \qty{32768}{\bitpersecond} for the address, utilizing a rectangular filter. 

\textsc{WakeMod}'s transmitter output power and spectrum were measured using an \textsc{R\&S FSIQ 3} signal analyzer. The signal analyzer was configured using \qty{50}{\deci\bel} internal attenuation, a reference level of \qty{20}{\deci\belmilliwatt}, and a resolution and video bandwidth of \qty{20}{\kilo\hertz}. The center frequency was \qty{868.324}{\mega\hertz} with a span of \qty{1}{\mega\hertz}. Measurements were performed by directly connecting a \textsc{WakeMod} soldered on a SMA-breakout board to the analyzer. To characterize the impact of system voltage on the transmit power, the spectrum was measured while varying the module's supply voltage from \qty{1.8}{\volt} to \qty{3.3}{\volt}.

\subsection{Packet Delivery Ratio Characterization}\label{subsec:pdr-setup}
The module's communication range was assessed outside in an open field. One \textsc{WakeMod} module was used as transmitter sending at \qty{2.8}{\deci\belmilliwatt}, while a second one was used as receiver -- both of them equipped with an omnidirectional dipole antenna with a gain of \qty{-2.1}{\deci\belisotrope} (\textsc{Linx ANT-868-CW-HWR-SMA}). The ground truth distances up to \qty{80}{\meter} were determined using a laser distance measurement device (\textsc{BOSCH GLM150-27C}), and using satellite photography up to \qty{150}{\meter}. The \ac{PDR} was characterized for different combinations of the low and high data rates. At each distance, a total of 100 \ac{WuC} was sent.

\begin{figure*}
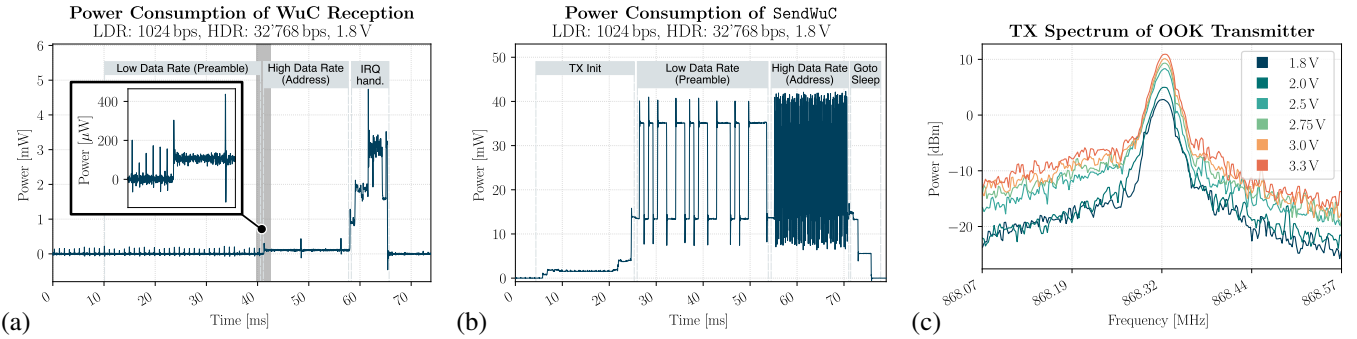

    \vspace{-0.5cm}
    \centering
    \stackinset{l}{0pt}{b}{0pt}{(a)}{
      \includesvg[width=0.31\textwidth]{figures/rx_ldr2_hdr7_edit.svg}
    }
    \stackinset{l}{0pt}{b}{0pt}{(b)}{
      \includesvg[width=0.31\textwidth]{figures/tx_1V8_ldr2_hdr7_edit.svg}
    }
    \stackinset{l}{0pt}{b}{0pt}{(c)}{
      \includesvg[width=0.31\textwidth]{figures/spectrum_edit.svg}
    }
    \vspace{-0.3cm}
    \caption{Power consumption of \textsc{WakeMod} during reception (a) and transmission (b) of a \ac{WuC}. (c) shows an analysis of the transmitter's spectrum at \qty{1.8}{\volt}.}\label{fig:pwr_spectrum_wakemod}
    \vspace{-0.3cm}
\end{figure*}

\subsection{WakeTag Demonstrator Evaluation}
In addition to characterizing the core module, \textsc{WakeTag} was used as a demonstration application, integrating \textsc{WakeMod} with an e-ink display. The overall power consumption of the complete \textsc{WakeTag} system was measured using the same \textsc{Keysight N6705C/N6781A} setup. The characterization focused on the power profile during reception of a \ac{WuC} with a payload of \qty{6}{\byte}, the rendering of the display, and the sleep consumption of the device.
For battery lifetime estimations, a \textsc{CR2032} with \qty{220}{\milli\ampere{}\hour} at a nominal voltage of \qty{3}{\volt} and an assumed 1\% annual self-discharge rate is considered. 

%% file: content/05_results.tex
\section{Results}\label{sec:results}
\subsection{Power Consumption and Transmitter Spectrum}
Measurements of the \ac{WuR} operation revealed that the power consumption during both idle listening and active reception phases is primarily determined by the configured data rate. No significant difference in power draw was observed between the module merely listening for a preamble and actively processing an incoming signal. As detailed in~\cref{tab:pwr}\,a), the consumption scales with the data rate, ranging from \qty{6.88}{\micro\watt} at \qty{1024}{\bitpersecond} up to \qty{105.88}{\micro\watt} at \qty{32768}{\bitpersecond}. Notably, these figures represent the total consumption of the \textsc{WakeMod} module, including its \ac{MCU}, \ac{OOK} transmitter, and antenna switch -- all operating in their deepest sleep state. This overall module consumption is approximately 9\% lower than the typical consumption figures specified in~\cite{frauenhofer_rficient} for the \textsc{FH101RF} itself. This data rate dependency allows for an energy-efficient strategy where a low data rate (e.g., \qty{1024}{\bitpersecond}) is used for the prolonged idle listening and preamble detection phase. Subsequently, the data rate can be increased for the shorter address decoding phase, reducing the active reception time and thereby minimizing latency and the required transmission duration for the sending device. For context, a preamble transmitted at \qty{1024}{\bitpersecond} requires \qty{31.25}{\milli\second}; while the address phase is 16 times longer in terms of bits, using a higher data rate significantly shortens its time duration. The selection of the \ac{LDR} and \ac{HDR} is thereby under the tradeoff between low-latency, \ac{WuC} frequency, and power consumption.

The power consumption during the transmission of a \ac{WuC} is presented in~\cref{tab:pwr}\,b) and~\cref{fig:pwr_spectrum_wakemod}\,b-c). A strong dependence on the supply voltage is evident. As the voltage increases from \qty{1.8}{\volt} to \qty{3.3}{\volt}, the transmitted power increases from \qty{2.78}{\deci\belmilliwatt} to \qty{10.92}{\deci\belmilliwatt}. Jointly, the module's power consumption during transmission rises substantially, from \qty{26.08}{\milli\watt} to \qty{108.54}{\milli\watt}. 
\cref{fig:pwr_spectrum_wakemod}\,b) shows a typical \ac{WuC} with the \ac{LDR} being \qty{1024}{\bitpersecond} and the \ac{HDR} being \qty{32768}{\bitpersecond}.

Finally, \cref{tab:pwr}\,c) quantifies the energy consumption and duration of auxiliary operations that do not involve continuous RX or TX. These include internal tasks performed by the \textsc{WakeMod}'s \ac{MCU}, such as handling an \ac{IRQ} from the \textsc{FH101RF} upon wake-up detection, and the overhead associated with communication between an external host \ac{MCU} and \textsc{WakeMod}. The baseline cost for the \textsc{WakeMod} \ac{MCU} to handle an \acs{IRQ} (without payload readout) is \qty{15.88}{\micro\joule} over \qty{7.4}{\milli\second}. Reading out an additional \qty{6}{\byte} payload during \acs{IRQ} handling increases the cost to \qty{46.64}{\micro\joule} over \qty{19.6}{\milli\second}. Communicating with the host incurs costs like \qty{26.59}{\micro\joule} for a \texttt{WhoAmI} command and \qty{1.14}{\milli\joule} for the \texttt{SetupWuR} operation (including \ac{WuR} calibration). As noted in the table, the energy cost listed for \texttt{SendWuC} (\qty{106.15}{\micro\joule}) represents only the command processing overhead and ramp up of transmitter; the significant energy consumption of the actual radio transmission must be added to determine the total energy cost of sending a \ac{WuC}.

Comparing two potential \ac{WuC} configurations highlights a key trade-off: Using a \qty{1024}{\bitpersecond} data rate to encode the preamble's \ac{mls}, and a \qty{32768}{\bitpersecond} data rate to encode the address' \ac{mls} results in a sender energy cost of \qty{1.33}{\milli\joule} (over \qty{72.58}{\milli\second}), while the receiver consumes \qty{17.75}{\micro\joule} during \qty{54.28}{\milli\second}. Crucially, the receiver's idle listening power in this mode is only \qty{6.88}{\micro\watt}. Conversely, using a faster \qty{32768}{\bitpersecond} preamble (and address) significantly reduces the sender's cost to \qty{539.12}{\micro\joule} in \qty{42.3}{\milli\second} and keeps a similar receiver's energy of \qty{17.64}{\micro\joule} during \qty{24.00}{\milli\second}. However, this requires the receiver to idle listen at a much higher \qty{105.88}{\micro\watt}. Therefore, while the faster preamble configuration is considerably more energy-efficient for the sender and faster for both parties during the actual wake-up transaction, the slower preamble configuration offers substantially lower idle power consumption for the receiver, leading to potentially much longer battery life in deployments where wake-up events are infrequent.

\subsection{Receiver Sensitivity and Packet Delivery Ratio}
The receiver's sensitivity was characterized to be \qty{-72.62}{\deci\belmilliwatt} with a peak envelope power of \qty{-70.18}{\deci\belmilliwatt}.

\begin{figure}[htpb!]
    \vspace{-0.3cm}
    \centering
    \includesvg[width=0.7\columnwidth]{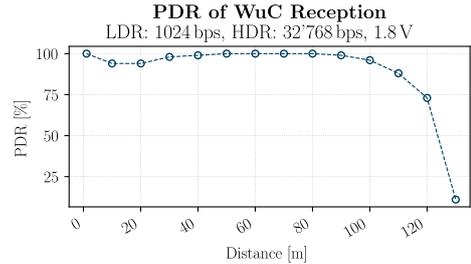}
    \vspace{-5mm}
    \caption{\Ac{PDR} of \aclp{WuC} sent and received by \textsc{WakeMod} across different distances.}
    \label{fig:pdr_wakemod}
\end{figure}

As presented in~\cref{fig:pdr_wakemod}, using the setup described in~\cref{{subsec:pdr-setup}}, a high wake-up reliability with \acp{PDR} above 94\% was observed for close ranges of \qty{1.1}{\meter} up to a distance of \qty{100}{\meter}. The \ac{PDR} starts then to exponentially decrease to 11\% at \qty{130}{\meter}. No \ac{WuC} were successfully received at distances above \qty{130}{\meter}, establishing the effective communication range limit under the tested conditions.

\begin{figure}[htpb!]
    \vspace{-0.3cm}
    \centering
    \includesvg[width=0.7\columnwidth]{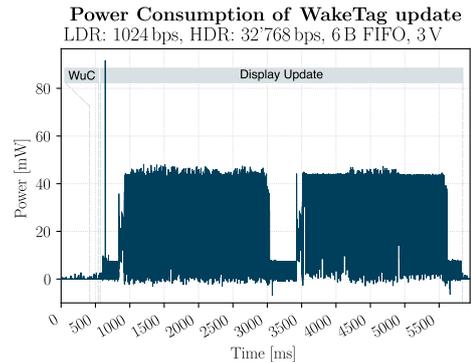}
    \vspace{-5mm}
    \caption{Power profile of \textsc{WakeTag} on receiving a single \ac{WuC} and refreshing the display.}
    \label{fig:pwr_display}
    \vspace{-0.5cm}
\end{figure}

\subsection{WakeTag Evaluation}
The power consumption of \textsc{WakeTag}, as illustrated in~\cref{fig:pwr_display}, is composed of three parts: idle-listening, the reception of the \ac{WuC}, and the update of the display.
With a consumption of \qty{7.17}{\micro\watt} in idle-listening, \textsc{WakeTag}'s consumption is only marginally higher than that of \textsc{WakeMod} (but includes voltage conversion and a second \ac{MCU}). As the update of the display costs \qty{132.22}{\milli\joule}, the reception of the \ac{WuC} itself is negligible.
The battery lifetime is therefore strongly correlating with the update frequency of the screen. High update frequencies of \qty{0.1}{\hertz} lead to a short estimated lifetime of 2 days, while an hourly update increases the lifetime to 1.7 years.
With daily updates, the \textsc{WakeTag} could theoretically operate for nearly \qty{8}{years}. The lifetime converges then to \qty{9.5}{years} when never an update of the display is initiated.

%% file: content/06_conclusion.tex
\section{Conclusion}\label{sec:conclusion}
In this paper, we present the design, the implementation, and a detailed evaluation of \textsc{WakeMod}, an open-source and ultra-low power wireless module leveraging a \ac{WuR} for energy-efficient, on-demand \ac{IoT} applications.

\textsc{WakeMod}'s capabilities were evaluated in extensive experiments, achieving an ultra-low idle-listening consumption of just \qty{6.9}{\micro\watt}, and an energy consumption per \ac{WuC} of \qty{17.7}{\micro\joule}. The corresponding wake-up latency was below \qty{54.3}{\milli\second}.
The integrated transmitter was fully characterized using a spectrum analyzer, achieving a peak transmission power of \qty{2.8}{\deci\belmilliwatt} at a power consumption of \qty{26.1}{\milli\watt}. Depending on the configuration of the \ac{WuR}, this enables the selective wake-up of remote \ac{IoT} devices with as little as \qty{539.1}{\micro\joule} and in less than \qty{42.3}{\milli\second}.
Further performance evaluations confirmed reliable communication with a high sensitivity of \qty{-72.6}{\deci\belmilliwatt}, achieving \acp{PDR} exceeding 94\% up to \qty{100}{\meter}, and an operational range limit of approximately \qty{130}{\meter}.

Finally, \textsc{WakeMod} was demonstrated in \textsc{WakeTag}, an e-ink price tag application showcasing the practical potential of \textsc{WakeMod}. Through careful system design, multi-year battery lifetimes of up to 8 years are achievable on a standard \textsc{CR2032} coin cell, while still allowing one display update per day. This illustrates the module's suitability for long-term, energy-constrained deployments where infrequent but low-latency communication is desired.